\documentclass[preprint]{sig-alternate-2013}

\usepackage{booktabs} % For formal tables

\permission{\copyright 2017 International World Wide Web Conference Committee (IW3C2),
published under Creative Commons CC BY 4.0 License.}
\conferenceinfo{WWW 2017,}{April 3--7, 2017, Perth, Australia.}
\copyrightetc{ACM \the\acmcopyr}
\crdata{978-1-4503-4914-7/17/04. \\
http://dx.doi.org/10.1145/3041021.3054213 \\
\includegraphics{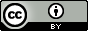}
}

\usepackage{graphicx}
\usepackage{array}
\usepackage{pbox}
\usepackage{wrapfig}
\usepackage{adjustbox}

\usepackage{amsmath,amssymb,amsfonts}
\usepackage{mathtools}

\usepackage{url}
\usepackage{verbatim}
\usepackage{multirow}
\usepackage{rotating}
\usepackage{todonotes}

\usepackage[linesnumbered,ruled,vlined]{algorithm2e}
\usepackage{algpseudocode}

\usepackage{ifpdf}
\usepackage{lmodern}
\usepackage{lipsum}
\usepackage[T1]{fontenc}
\usepackage{textcomp}
\let\labelindent\relax

\usepackage{enumitem}
\usepackage{mdwlist}

\usepackage{listings}

\usepackage{comment}
\usepackage{ifthen}
\usepackage{color}
\usepackage{colortbl,xcolor}

\usepackage{hhline}

\definecolor{LightCyan}{rgb}{0.88,1,1}
\definecolor{Gray}{gray}{0.9}
\definecolor{lightgray}{rgb}{0.83, 0.83, 0.83}
\definecolor{darkgray}{rgb}{0.66, 0.66, 0.66}
\specialcomment{notes}{\begingroup \color{blue}} { \endgroup}

\colorlet{punct}{red!60!black}
\definecolor{background}{HTML}{EEEEEE}
\definecolor{delim}{RGB}{20,105,176}
\colorlet{numb}{magenta!60!black}

\makeatletter
\newcommand*{\inlineequation}[2][]{%
  \begingroup
    \refstepcounter{equation}%
    \ifx\\#1\\%
    \else
      \label{#1}%
    \fi
    \relpenalty=10000 %
    \binoppenalty=10000 %
    \ensuremath{%
      #2%
    }%
    ~\@eqnnum
  \endgroup
}
\makeatother

\lstdefinelanguage{json}{
    basicstyle=\small\ttfamily,
    numbers=left,
    numbers=none,
    stepnumber=1,
    numbersep=8pt,
    showstringspaces=false,
    breaklines=true,
    frame=lines,
    backgroundcolor=\color{background},
    literate=
     *{0}{{{\color{numb}0}}}{1}
      {1}{{{\color{numb}1}}}{1}
      {2}{{{\color{numb}2}}}{1}
      {3}{{{\color{numb}3}}}{1}
      {4}{{{\color{numb}4}}}{1}
      {5}{{{\color{numb}5}}}{1}
      {6}{{{\color{numb}6}}}{1}
      {7}{{{\color{numb}7}}}{1}
      {8}{{{\color{numb}8}}}{1}
      {9}{{{\color{numb}9}}}{1}
      {:}{{{\color{punct}{:}}}}{1}
      {,}{{{\color{punct}{,}}}}{1}
      {\{}{{{\color{delim}{\{}}}}{1}
      {\}}{{{\color{delim}{\}}}}}{1}
      {[}{{{\color{delim}{[}}}}{1}
      {]}{{{\color{delim}{]}}}}{1},
}

\newcolumntype{!}{>{\global\let\currentrowstyle\relax}}
\newcolumntype{^}{>{\currentrowstyle}}

\DeclareGraphicsExtensions{.eps,.pdf,.png,.jpeg}
\graphicspath{{figure/eps/},{figure/pdf/},{figure/png/},{figure/jpeg/}}

\newcommand{\superscript}[1]{\ensuremath{^{\textrm{#1}}}}

\newcommand{\si}{\begin{enumerate}[leftmargin=*, itemindent=0cm, align=left]\itemsep0em}

\newcommand{\ei}{\end{enumerate}}

\toappear{WWW 2017 Companion, April 3-7, 2017, Perth, Australia.}
\begin{document}

\title{Global Entity Ranking Across Multiple Languages}

\def\kloutinc{\superscript{*}}

\numberofauthors{1}
\author{
   \alignauthor Prantik Bhattacharyya, Nemanja Spasojevic \\
   \affaddr{Lithium Technologies | Klout}\\
   \affaddr{San Francisco, CA}\\
   \email{\{prantik.bhattacharyya, nemanja.spasojevic\}@lithium.com}
}

\maketitle

\begin{abstract}
We present work on building a global long-tailed ranking of entities across multiple languages using Wikipedia and Freebase knowledge bases.
We identify multiple features and build a model to rank entities using a ground-truth dataset of more than $10$ thousand labels.
The final system ranks $27$ million entities with $75\%$ precision and $48\%$ F1 score.
We provide performance evaluation and empirical evidence of the quality of ranking across languages, and open the final ranked lists for future research.
\end{abstract}

% \category{TBD}{TBD}

\nocite{}
\vspace{-0.1in}
\keywords{entity ranking; entity extraction; knowledge base;}

\vspace{-0.08in}
\section{Introduction}
\label{section:introduction}
In the past decade, a number of openly available Knowledge Bases (KBs) have emerged.
The most popular ones include Freebase, Wikipedia, and Yago, containing around 48M, 25M, and 10M entities respectively.
Many of the entities overlap across the KBs.
In NLP entity linking \footnote{also known as named entity linking (NEL),
named entity disambiguation (NED) or named entity recognition and disambiguation (NERD)},
the task is to link mentioned entities within text to their identity within the KB.
A foundational part of setting up a real-time entity linking system is to choose which entities to consider,
as memory constraints prohibit considering the entire knowledge base \cite{Bhargava:edl}.
Additionally, some entities may not be of relevance. %to consider either.
In order to maximize quality of the NLP entity linking system, we need to include as many important entities as possible.

In this paper we identify a collection of features to perform scoring and ranking of the entities.
We also introduce the ground truth data set that we use to train and apply the ranking function.

\begin{comment}
select language, count(*) from wiki_link_features where dt = '${wikiFeatureDate}' group by language order by language

ar	1257616
ca	629859
de	2542175
en	9127360
es	2816547
fa	2337295
fi	669319
fr	2474320
hi	158227
id	828822
it	2200671
ja	1634914
ko	846929
nl	2455348
no	867804
pl	1685790
pt	1647817
ru	2685176
sr	790911
sv	3717193
uk	1250394
vi	1533406
zh	2408647
zh_yue	71241

select language, count(*) from entity_features where dt = 'LATEST_DEV' group by language order by language

ar	416220
ca	296610
de	833265
en	4,316,567
es	704126
fa	340347
fi	498496
fr	953333
hi	76414
id	296446
it	756930
ja	408698
ko	398388
nl	675142
no	526516
pl	661262
pt	564328
ru	586104
sr	390620
sv	523841
uk	604054
vi	774798
zh_cn	426837
~	43,796,610

\end{comment}
\vspace{-0.1in}

\section{Related Work}
\label{section:related_work}
A large body of previous work has addressed ranking entities in terms of temporal popularity, as well as in the context of a query; however, little study has been done in terms of building the global rank of entities within a KB.
Temporal entity importance on Twitter was studied by Pedro et.~al.~\cite{saleiro2016learning}.
In \cite{gionis2012estimating}, authors propose a hybrid model of entity ranking and selection in the context of displaying the most important entities for a given constraint while eliminating redundant entities.
Entity ranking of Wikipedia entities in the context of a query, has been done using link structure and categories \cite{vercoustre2008entity}, as well as graph methods and web search \cite{zaragoza2007ranking}.

%Wang et.~al.\cite{wang2016nerank} studied problem of ranking within a document without of query context where they used graphical model (Topical Tripartite Graph) to derive ranking.
\vspace{-0.1in}

\section{Our Approach}
\label{section:problem_definition}
Given KB, we want to build a \textbf{global long-tailed ranking} of entities in order of socially recognizable importance.
When building the NLP entity linking system, the $N$ top ranked entities from KB should yield maximum perceived quality by casual observers.
\vspace{-0.05in}
\subsection{Data Set}
\label{subsection:data_set}

We collected a labeled data set by selecting $10,969$ entities.
We randomly sampled as well as added some important entities, to balance the skewed ratio that KBs have of important / non-important entries.
Each evaluator had to score the entities on scale 1 to 5; 5 being most important.
Seven evaluators used the following guidelines regarding importance:

\begin{description}[noitemsep, nolistsep, style=unboxed,leftmargin=0.5cm]
  \item [Public Persons] important if currently major pro athletes, serving politicians, etc.
           If no longer active, important if influential (e.g. Muhammad Ali, Tony Blair).
  \item [Locations] look at population (e.g. Albany, California vs. Toronto, Canada), historical significance (Waterloo).
  \item [Dates] unimportant unless shorthand for a holiday or event (4th of July, 9/11).
  \item [Newspapers] important, especially high-circulation ones (WSJ).
  \item [Sports Teams] important if in pro league.
  \item [Schools] important if recognised globally.
  \item [Films \& Song] major franchises and influential classics are important -- more obscure are often not.
  \item [Laws] important if they enacted social change (Loving v. Virginia, Roe v. Wade), unimportant otherwise.
  \item [Disambiguators] entities that disambiguate are important because we want them in the dictionary (Apple, Inc.~and Apple Fruit).
\end{description}

% unimportant  <----> important
% # of entries
% # break down by type
% # number of evaluators
%
% 1-5 , 5 - extremaply important ()
%
% 1- important
% 2442 entries

\subsection{Features and Scoring}
\label{subsection:rankin}

Features were derived from Freebase and Wikipedia sources.
They capture popularity within Wikipedia links, and how important an entity is within Freebase.
Some signals used are page rank, link in/out counts and ratio, number of categories a page belongs to in Wikipedia.
We also use the number of objects, a given entity is connected to, i.e., object and object type count features, as well as %type descriptors within knowledge base
the number of times a given entity was an object with respect to another entity, i.e., subject and subject type count features.
We also extract social media identities mentioned in an entity's KB and use their Klout score \cite{rao2015kloutScore} as a feature.
The full set of features derived as well as their performance is listed in Table \ref{table:feature_prec}.

We model the evaluator's score using simple linear regression.
The feature vector $\mathcal{F}(e)$ for an entity $e$ is represented as:
$\mathcal{F}(e) = [f_1(e), f_2(e), ..., f_m(e)]
$ where $f_k(e)$ is the feature value associated with a specific feature $f_k$.
Normalized feature values are denoted by $\hat{f_{k}}(e)$.
Features are normalized as:
$\hat{f_{k}}(e) = \frac {log(f_{k}(e))}{\operatorname*{max}\limits_{e_i \in KB} log(f_{k}(e_i))}$.
Importance score for an entity is denoted by $\mathcal{S}(e)$ and is computed as the dot product of a
weight vector $\mathbf{w}$ and the normalized feature vector:
% $\mathcal{S}(e) = \mathbf{w} \cdot \hat{\mathcal{F}}(e) \label{equation:score}$
\inlineequation[equation:score]{\mathcal{S}(e) = \mathbf{w} \cdot \hat{\mathcal{F}}(e)}.
Weight vector is computed with supervised learning techniques, using labeled ground truth data (train/test split of 80/20).

\vspace{-0.1in}

\section{Experiments}
\label{section:problem_definition}
% Final score:
% select *
% from entity_importance_metric
% where dt = '20161220' and
%       source = 'ENTITY_IMP_PREC_REC'

\begin{table}
\small
\caption{Feature Performance For English Rankings}
\vspace{-0.14in}
\begin{center}
\resizebox{\columnwidth}{!}{
\begin{tabular}{|l|l|c|c|c|c|c|}
\hline
\rowcolor{Gray}
 \multicolumn{2}{|c|}{\textbf{Feature}} & \textbf{P} & \textbf{R} & \textbf{F1} & \textbf{C} & \textbf{RMSE} \\ \hline
  \parbox[t]{2mm}{\multirow{6}{*}{\rotatebox[origin=l]{90}{ Wikipedia}}}
  & \cellcolor[rgb]{0.859,0.859,0.859}Page Rank & 0.59 & 0.05 & 0.09 & 0.164 & 1.54 \\ \cline{2-7}
  & \cellcolor[rgb]{0.859,0.859,0.859}Outlink Count & 0.55 & 0.13 & 0.21 & 0.164 & 2.09 \\ \cline{2-7}
  & \cellcolor[rgb]{0.859,0.859,0.859}Inlink Count & 0.62 & 0.12 & 0.20 & 0.164 & 1.82 \\ \cline{2-7}
  & \cellcolor[rgb]{0.859,0.859,0.859}In Out Ratio & 0.75 & 0.19 & 0.31 & 0.164 & 1.54 \\ \cline{2-7}
  & \cellcolor[rgb]{0.859,0.859,0.859}Category Count & 0.65 & 0.21 & 0.36 & 0.164 & 1.89 \\ \hline
  \parbox[t]{2mm}{\multirow{3}{*}{\rotatebox[origin=l]{90}{Freebase }}}
  & \cellcolor[rgb]{0.859,0.859,0.859}Subject \# & 0.28 & 0.06 & 0.10 & 1.000 & 2.39 \\ \cline{2-7}
  & \cellcolor[rgb]{0.859,0.859,0.859}Subject Types \# & 0.42 & 0.10 & 0.16 & 1.000 & 2.25 \\ \cline{2-7}
  & \cellcolor[rgb]{0.859,0.859,0.859}Object \# & 0.62 & 0.12 & 0.20 & 0.973 & 2.00 \\ \cline{2-7}
  & \cellcolor[rgb]{0.859,0.859,0.859}Object Types \# & 0.46 & 0.11 & 0.17 & 0.973 & 2.25 \\ \hline
  & \cellcolor[rgb]{0.859,0.859,0.859}Klout Score & 0.57 & 0.11 & 0.17 & 0.004 & 2.32 \\ \hline
  & \cellcolor[rgb]{0.859,0.859,0.859}$\mathcal{S}(e)$ - All Feat. & \textbf{0.75} & \textbf{0.37} & \textbf{0.48} & \textbf{1.00} & \textbf{1.15} \\ \hline
\end{tabular}
}
\end{center}
\label{table:feature_prec}
\vspace{-0.18in}
\end{table}

%{"prod.text.entity_importance.en.weighted_f1":0.5019177210104604,"prod.text.entity_importance.en.weighted_recall":0.37572632785006954,"prod.text.entity_importance.en.weighted_precision":0.7557406838952564}
%Time taken: 64.149 seconds, Fetched: 1 row(s)
%hive>
%    >
%    >
%    > select * from tmp_entity_imp_raw_prec_recall;
%select * from tmp_entity_imp_raw_prec_recall
%OK
%0	3	1130	11086	0.0	0.0	0.0	1130	1
%18	47	957	11197	0.27692307692307694	0.018461538461538463	0.03461538461538462	975	2
%218	1909	298	9794	0.10249177244945933	0.42248062015503873	0.16496405599697314	516	3
%120	5463	141	6495	0.021493820526598602	0.45977011494252873	0.04106776180698152	261	4
%4235	205	5102	2677	0.9538288288288288	0.45357181107422084	0.6147927705596283	9337	5
%

\begin{figure}
 \centering
 \includegraphics[height=0.45\linewidth]{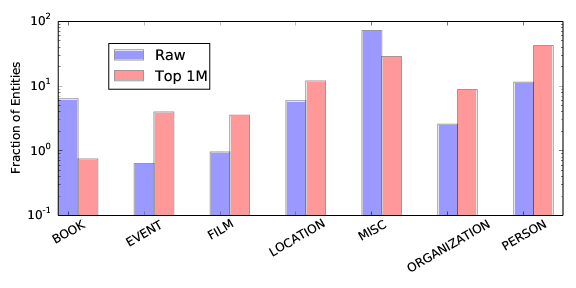}
 \vspace{-0.16in}
 \caption{Entity Count by Type}
 \label{fig:entity_count_by_type}
 \vspace{-0.2in}
\end{figure}

Table \ref{table:feature_prec} shows precision, recall, F1 and the population coverage for the full list of features and the final system.
The importance score was calculated using Eq.\ref{equation:score} where final score was rounded to an integer value
so it can be compared against the labels from ground-truth data.

We observe that Wikipedia features have the highest precision among all the features.
The Freebase features have the highest coverage values.
The Klout score feature also has one of the highest individual precision values.
While this feature has the lowest coverage, it helps boost the final score and floats up a few relevant entities for final system application in social media platforms.
We also look at root mean squared error (RMSE) of the entity scores against assigned labels.
The final model shows the lowest RMSE value.

We also plot the distribution of entity types in the top $1$ million ranked entities and the unranked list for the English language.
$11\%$ of entities are of type `person' in the global list while the top ranked list contains $42\%$ entities of type `person'.
The percentage of `MISC' entity types drop from $72\%$ to $29\%$.
%$12\%$ of the entities of type `person' are ranked in the top $1$ million compared to only $0.39\%$ of entities of the type `book'.
These difference in coverage highlight that entities are ranked relevantly in the corpus. %the list can successfully rank capture relevant entities from a large corpus of entities.

\begin{table}
  \small
  \centering
  \begin{tabular}{ |@{\hskip1pt}c@{\hskip1pt}|@{\hskip3pt}c@{\hskip3pt}|@{\hskip2pt}c@{\hskip2pt}|@{\hskip2pt}c@{\hskip2pt}|@{\hskip2pt}c@{\hskip2pt}|@{\hskip2pt}c@{\hskip2pt}|@{\hskip2pt}c@{\hskip2pt}| }
    \hline
    \textbf{Entity} & \textbf{Image} & \textbf{EN} & \textbf{AR} & \textbf{ES} & \textbf{FR} & \textbf{IT}
%    \textbf{Entity} & \textbf{Image} & \textbf{English} & \textbf{Arabic} & \textbf{Spanish} & \textbf{French} & \textbf{Italian}

    \\ \hline
    Vogue &
    % 01y5zy &
    \begin{minipage}{.05\textwidth} \includegraphics[width=\linewidth]{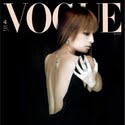}\end{minipage}
    & 2 & 6,173 & 200 & 2,341 & 62

    \\ \hline
    \pbox{20cm}{World \\ Bank} &
    % 02vk52z &
    \begin{minipage}{.05\textwidth} \includegraphics[width=\linewidth]{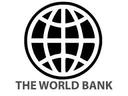}\end{minipage}
    & 322 & 103 & 3,747 & 2,758 & 5,704

    \\ \hline
    Morocco &
    % 04wgh &
    \begin{minipage}{.05\textwidth} \includegraphics[width=\linewidth]{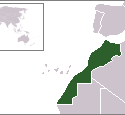}\end{minipage}
    & 1,277 & 2 & 527 & 544 & 232

    \\ \hline
    \pbox{10cm}{Donald \\ Duck} &
    % 02gln &
    \begin{minipage}{.05\textwidth} \includegraphics[width=\linewidth]{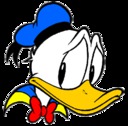}\end{minipage}
    & 10,001 & 9,494 & 7,444 & 10,380 & 4,575

    \\ \hline
    Balkans &
    % 01jjk &
    \begin{minipage}{.05\textwidth} \includegraphics[width=\linewidth]{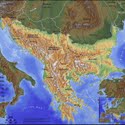}\end{minipage}
    & 36,753 & 109 & 17,456 & 9,383 & 2,854

%    \\ \hline
%    Fidel Castro &
%    % 09k0f &
%    \begin{minipage}{.05\textwidth} \includegraphics[width=\linewidth]{freebase_image/09k0f}\end{minipage}
%    & 6,096 & 4,999 & 8,579 & 13,886 & 2,369

%    \\ \hline
%    POSIX &
%    % 05t0l &
%    \begin{minipage}{.05\textwidth} \includegraphics[width=\linewidth]{freebase_image/05t0l}\end{minipage}
%    & 54,040 & 100,003 & 48,235 & 46,810 & 15,030

    \\ \hline
    Bed &
    % 03ssj5 &
    \begin{minipage}{.05\textwidth} \includegraphics[width=\linewidth]{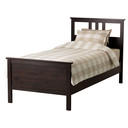}\end{minipage}
    & 109,686 & 23,809 & 68,180 & 66,859 & 52,713

    \\ \hline
    \pbox{20cm}{Bunk \\ Bed} &
    % 03f982 &
    \begin{minipage}{.05\textwidth} \includegraphics[width=\linewidth]{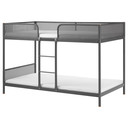}\end{minipage}
    & 992,576 & 64,399 & 330,669 & 906,988 & 416,292

    \\ \hline

  \end{tabular}
  \vspace{-0.1in}
  \caption{Entity Ranking Examples For Different Languages}
  \label{table:rank_examples}
  \vspace{-0.2in}
\end{table}

% select language, rank, machine_id, debug_entity(machine_id), raw_importance
% from entity_importance
% where ((rank > 0 and rank < 10) or
%        (rank > 100 and rank < 110) or
%        (rank > 1000 and rank < 1010) or
%        (rank > 10000 and rank < 10010) or
%        (rank > 100000 and rank < 100010) or
%        (rank > 1000000 and rank < 1000010)) and
%       dt = 'LATEST_PROD' and
%       source = 'REGRESSION' and
%       language in ('en', 'es', 'ar')
% order by language, rank, machine_id
% limit 1000
% ;
%
%
% select language, rank, machine_id, debug_entity(machine_id)
% from entity_importance
% where machine_id in
%       ( '01y5zy',  -- Vogue
%         '04wgh',   -- Morocco
%         '02vk52z', -- WorldBank
%         '05t0l',   -- POSIX
%         '01jjk',   -- Balkan
%         '02gln',   -- Donald Duck
%         '03f982',  -- Bunk Beds
%         '0fv6x',   -- Laibach
%         '09k0f',   -- Fidel
%         '03ssj5'   -- Bed
%       ) and
%       dt = 'LATEST_PROD' and
%       source = 'REGRESSION' and
%       language in ('en', 'es', 'ar', 'fr', 'it')
% order by language, rank, machine_id
% limit 1000
% ;

In Table \ref{table:rank_examples}, we provide examples of entities with their ranks in a particular language.
We see that the entity ranks are regionally sensitive in the context of their language, e.~g. `Morocco' is ranked $2$ in the ranking for `Arabic' language.
%linguistic sensitivity
We also observe the rankings are sensitive with respect to the specificity of the entity, for example `bunk bed' is ranked magnitudally lower than the more generic entity `bed'.
%`Vogue` has high ranks for all languages except Arabic.

\vspace{-0.05in}
\
\section{Summary}
\label{section:conclusion}
We make the ranked list of top $500,000$ entities available as an open source data set at \url{https://github.com/klout/opendata}.
To conclude, in this work, we built a global ranking of entities across multiple languages combining features from multiple knowledge bases.
We also found that combination of multiple features yields the best results.
Future work in this direction is to include new signals such as Wikipedia page view statistics and edit history.


{\small
\bibliographystyle{abbrv}
\bibliography{bibliography}
}

% ACM needs 'a single self-contained file'!

% \section{Appendix}
% \label{section:appendix}
% \input{texfiles/experts_appendix}

% \balancecolumns
% That's all folks!
\end{document}